\documentclass[aip,jmp,preprint,amsmath,amssymb,showpacs]{revtex4-1}
\usepackage{amssymb}
\usepackage{epsfig}
\usepackage{amsmath}
\usepackage{times}

\begin{document}

\title{A model of spreading of sudden events on social networks }
\author{Jiao Wu}
\affiliation{Business School, University of Shanghai for Science and Technology, Shanghai 200093, China}
\author{Muhua Zheng}
\email{zhengmuhua163@gmail.com}
\affiliation{Department of Physics, East China Normal University, Shanghai, 200062, China}
\author{Zi-Ke Zhang}
\affiliation{Alibaba Research Center for Complexity Sciences, Hangzhou Normal University, Hangzhou 311121, China}
\author{Wei Wang}
\affiliation{College of Computer Science and Technology, Chongqing University of Posts and Telecommunications,
Chongqing 400065, China}
\author{Changgui Gu}
\email{gu\_changgui@163.com}
\affiliation{Business School, University of Shanghai for Science and Technology, Shanghai 200093, China}
\author{Zonghua Liu}
\affiliation{Department of Physics, East China Normal University, Shanghai, 200062, China}

\begin{abstract}
Information spreading has been studied for decades, but its underlying mechanism is still under debate,
especially for those ones spreading extremely fast through Internet. By focusing on the information
spreading data of six typical events on Sina Weibo, we surprisingly find that the spreading of modern
information shows some new features, i.e. either extremely fast or slow, depending on the individual
events. To understand its mechanism, we present a Susceptible-Accepted-Recovered (SAR) model with both
information sensitivity and social reinforcement. Numerical simulations show that the model can reproduce
the main spreading patterns of the six typical events. By this model we further reveal that the spreading
can be speeded up by increasing either the strength of information sensitivity or social reinforcement.
Depending on the transmission probability and information sensitivity, the final accepted size can change
from continuous to discontinuous transition when the strength of the social reinforcement is large.
Moreover, an edge-based compartmental theory is presented to explain the numerical results. These findings may be of
significance on the control of information spreading in modern society.
\end{abstract}

\maketitle

\begin{quotation}
In modern society, our life depends more and more on the Internet and its related services such as live chat,
navigation, and online shopping etc. Consequently, some new forms of information spreading have emerged from
time to time such as the Facebook, Twitter, Weibo etc, which result in some new features of communication
activities. Thus, how to understand these new features of information spreading in social networks is a new
challenging problem. We here investigate the spreading phenomena of six typical events on Sina Weibo data and
surprisingly find that the spreading patterns may show very distinctive behaviours, i.e. some are extremely
fast while others slow, depending on the individual events. To understand them, we present a model to study
these new features. Based on this model, we reveal that there are two key factors to information spreading,
i.e. the information sensitivity and social reinforcement. Moreover, we find that the final spreading range
influenced by these two factors exhibits a discontinuous transition when the strength of the social
reinforcement is large. These findings open a new window to study the new features caused by the modern
communication tools.
\end{quotation}

\section{Introduction}
Information spreading on complex networks has been well studied and a lot of great progresses have been achieved \cite{Satorras:2015,Young:2011,Ratkiewicz:2010,Sornette:2004,Barrat:2008,Wang:2017,Liu:2015,Zhang:2016},
such as in the aspects of spreading patterns \cite{Nematzadeh:2014,Nagata:2014}, spreading threshold
\cite{Wang:2015,Zhu:2017}, propagation paths \cite{Prapivsky:2011}, human activities patterns
\cite{Watts:1998,Perez:2011} and source locating \cite{Santos:2005,Maslov:2002} etc. Recently, the attention has
been moved to the mechanism of explosive spreading \cite{Gardenes:2016,LiuQ:2017}, which corresponds to a
discontinuous transition in the phase space and may explain why the information can be accepted by many people
overnight. It has been revealed that the explosive spreading can be induced by incorporating some key properties
into the dynamics, such as the synergistic effects \cite{Gardenes:2016,LiuQ:2017}, social reinforcement
\cite{Wang:2015a,Guilbeault:2017}, threshold model \cite{Granovetter:1978,Watts:2002,Centola:2007}, memory effects
\cite{Dodds:2004,Janssen:2004,Bizhani:2012,Chung:2014}, non-linear cooperation of the transmitting spreaders
\cite{Liu:1987,Assis:2009}, and adaptive rewiring \cite{Gross:2006} etc. These results significantly increase our
understanding on the mechanism why ideas, rumours or products can suddenly catch on\cite{Gardenes:2016} and are
also very useful for us to create viral marketing campaigns, block the rumor spreading, evaluate the quality of
information, and predict how far it will spread.

Although many significant properties of information spreading have been uncovered in the previous studies, the
study on the individual's attitude towards an event or information (i.e., the information sensitivity) is neglected.
In addition, to the best of our knowledge, most of the previous studies mainly focus on how the spreading is
influenced by the network structure and other concrete factors through theoretical models, which lacks the supports
of real data. Therefore, it is necessary to investigate the effect of information sensitivity in information
spreading through both real data and theoretical models. For this purpose, we have tracked some spreading processes
of typical events on the largest micro-blogging system in China \cite{Liu:2015}$-$Sina Weibo (http://weibo.com/) and
obtained some historical spreading data about the corresponding events. Analyzing these data, we find that for the
sensitive events, their information is born with fashionable features and spreads rapidly; while for the insensitive
events, their information is doomed to be out of the public attention and spreads slowly. These findings call our
great interest and motivate us to study their underlying mechanism. As we know, the individual's attitude and interest
to different events are usually different, which may lead to very different performances
\cite{Crane:2008,Sano:2013,Wu:2007}. Thus, the information sensitivity to each specific event should be paid more
attention. In addition, Centola's experiments on behavior spreading \cite{Centola:2010} showed that the social
reinforcement (i.e., an individual requires multiple prompts from neighbors before adopting an opinion or behavior
\cite{Zheng:2013,Majumdar:2001,Young:2009,Kerchove:2009,Karrer:2010,Onnela:2010,Lu:2011})
typically play an important role in the adoption of information or behavior, which is an essential property of
information spreading. In this sense, we believe that it is very necessary to incorporate both the information
sensitivity and social reinforcement into the spreading dynamics.

In this work, we propose a model to emphasize the effects of both the information sensitivity (i.e., the primary
accepted probability in the model) and social reinforcement. Our numerical simulations reveal that the model can
show the main features of the six typical events obtained on Sina Weibo. Moreover, we find that increasing the
strength of information sensitivity and social reinforcement can significantly promote the information spreading.
We interestingly show that the final accepted size influenced by the transmission probability and information
sensitivity exhibits an abrupt increase, provided that the social reinforcement is large. An edge-based compartmental
theory is presented to explain the numerical results.

The rest of this paper is organized as follows. In Sec. II, the spreading data of six typical events on Sina Weibo
is collected and analyzed. In Sec. III, a model is presented to study the effects of information sensitivity
and social reinforcement. In Sec. IV, simulation results are presented and the effects of information sensitivity,
social reinforcement and transmission probability are discussed. In Sec. V, an edge-based compartmental theory is
given to explain the numerical results. Finally, in Sec. VI, the conclusions and discussions are presented.

\section{Data description}
To study the spreading of sudden events on social networks, we firstly extract some typical data from the Internet.
In detail, we choose the Sina Weibo (http://weibo.com/) as our source of data, which is one of the largest
micro-blogging system in China \cite{Liu:2015} and evolves about $20\%$ of the Chinese population. When an event
occurs, individuals usually post short messages to talk about it in online social network, namely tweets in Weibo.
At the same time, the follower individuals can follow their neighbors to forward the message (i.e., retweet), which
is very similar to Twitter. Thus, the Sina Weibo can efficiently reflect the spreading tendency of different events
and the sensitive intensity to the event. We have tracked some historical events spread in Sina Weibo from September
2009 to February 2012 and chosen those events with diversity so that to represent a wide-range of topics. As a
consequence, we find six typical events related to various aspects of social life, including public figures, natural
disaster, traffic accident, and so on. Table \ref{table1} shows the basic statistics of the six typical events. The
details of data of these six events are as follows \cite{Liu:2015}:

\textbf{(a) Wenzhou Train Collision}: Two high-speed trains (TVG) travelling on the Yongtaiwen railway
line collided at a viaduct in the suburb of Wenzhou in Zhejiang province.

\textbf{(b) Yushu Earthquake}: Yushu County, located on the Tibetan plateau in China, was awoken by a
magnitude 6.9 earthquake.

\textbf{(c) Death of Wang Yue}: (also referred to as the ¡ùxiao yueyue¡ì event) Wang Yue, a two year old
Chinese girl, was killed in a car crash by two vehicles in a narrow street in Foshan city, Guangdong
province.

\textbf{(d) Case of Running Fast Car in Hebei University}: Two students were hit by a car driven by a
drunk man at a narrow lane inside the Hebei University in Hebei province.

\textbf{(e) Tang Jun Education Qualification Fake}: Tang Jun, the well-known and successful former
president of Microsoft China and Shanda Interactive Entertainment, was accused by Fang Zhouzi, a
crusader against scientific and academic fraud, of falsifying his academic credentials and also patents.

\textbf{(f) Yao Ming Retire}: Yao Ming officially announced his retirement from basketball after nine
seasons in the team Houston Rockets.

\newcommand{\tabincell}[2]{\begin{tabular}{@{}#1@{}}#2\end{tabular}}
\begin{table}[b] \small
\caption{Basic statistics of the six typical events. Date indicates the time of corresponding event,
$N_m(100)$ represents the cumulative number of messages talking about the corresponding event till 100 days, $\Delta$
denotes the incremental rate of messages posted within the first 10 days.}\label{table1}
\begin{center}
\begin{tabular}{ccccc}
\\[-0.5cm]
\toprule
No. & Events & Date & $N_m(100)$ & $\Delta$ \\
\hline
a & Wenzhou Train Collision &  23/Jul./2011 & 281 905 &  0.9038\\

b & Yushu Earthquake &  14/Apr./2010 & 24 544 & 0.8771\\

c & Death of Wang Yue &  13/Oct./2011 & 148 297 & 0.7029\\

d &  \tabincell{c}{Case of Running Fast Car \\ in Hebei University} &  16/Oct./2010 & 74 156 & 0.1296 \\

e &  \tabincell{c}{Tang Jun Educatioin \\ Qualification Fake} &  01/Jul./2010 & 6776 & 0.2473\\

f &  Yao Ming Retire &  20/Jul./2011 & 45 006 &  0.3015\\
\toprule
\end{tabular}
\end{center}
\end{table}

\begin{figure}
\epsfig{figure=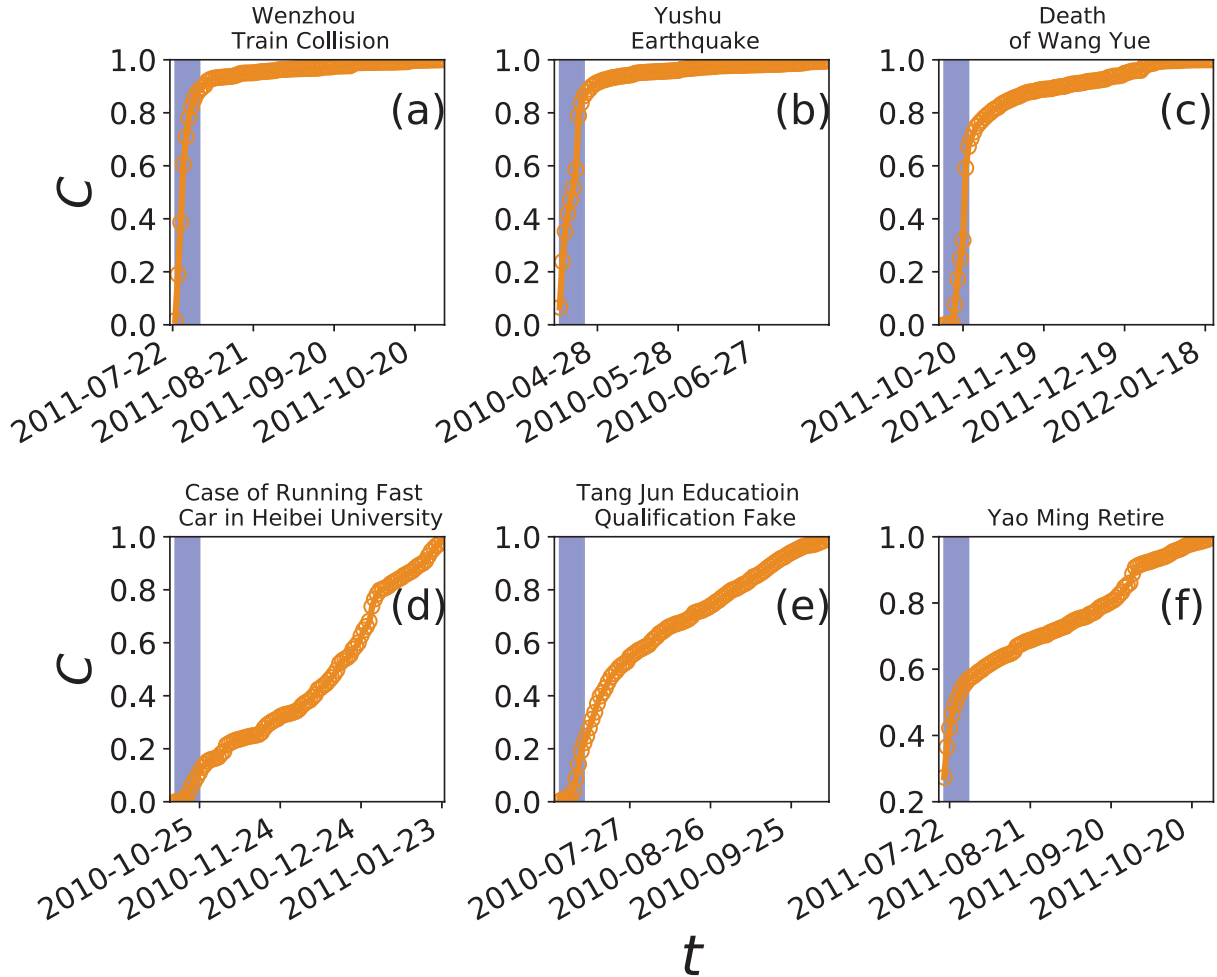,width=1.0\linewidth} \caption{(color
online). The spreading dynamics versus time $t$ for the six typical events on Sina Weibo, where (a)-(f)
represent the events of Wenzhou Train Collision, Yushu Earthquake, Death of Wang Yue, Case of Running Fast
Car in Hebei University, Tang Jun Educatioin Qualification Fake, and Yao Ming Retire, respectively. Light
blue areas represent the spreading range within the first $10$ days of the events.}
\label{Fig:data}
\end{figure}

To quantitatively describe the spreading dynamics of the six selected events, we define $C=\frac{N_m(t)}{N_m(100)}$
as the cumulative probability of messages talking about a specific event, where $N_m(t)$ and $N_m(100)$ represent
the cumulative number of messages posted until time $t$ and the $100th$ day, respectively. As individuals'
attention to an event decays very fast \cite{Wu:2007}, the information spreading process can be regarded as ended
after $100$ days. Fig. \ref{Fig:data}(a)-(f) show the spreading patterns of the six selected events within the
first $100$ days, respectively. From Fig. \ref{Fig:data}(a)-(c) we observe that these three events spread rapidly
in the first $10$ days (shaped by Light blue), indicating that people are very sensitive to these kinds of events.
A common point of Fig. \ref{Fig:data}(a)-(c) is that they represent the events of natural disaster or traffic
accident. In contrast, from Fig. \ref{Fig:data}(d)-(f) we observe that they spread slowly in both the first $10$
days and the subsequent days, indicating that people are insensitive to these kinds of events. A common point of
Fig. \ref{Fig:data}(d)-(f) is that they are human related events and thus are out of public attention. To better
characterize the difference of these spreading patterns, similar to Ref \cite{Liu:2015}, we here regard the time
window of the first $10$ days as the early stage. Then, we let $\Delta$ be the incremental rate of messages posted
within the first $10$ days, defined as $\Delta=\frac{N_m(10)-N_m(0)}{N_m(100)}=C(10)-C(0)$. In general, $C(0)$ is
zero when the event occurs except for the event of ``Yao Ming Retire", where some messages have been posted in
Weibo before the event occurs as someone has obtained the gossip of Yao's event from different channels \cite{Liu:2015}.
Further, we calculate the values of $\Delta$ for different events. Very interestingly, we find that for the
sensitive events, the information is of fashionable features and the values of $\Delta$ are very large. While for
the insensitive events, the information is dull for public attention with a relatively small $\Delta$. For example,
the event of Wenzhou Train Collision (Fig. \ref{Fig:data}(a)) is very attractive and its $\Delta$ can reach
$90.38\%$ within the first $10$ days. While for the event of Running Fast Car in Hebei University, it spreads
slowly and the $\Delta$ is just $12.96\%$ in the early stage.

\begin{figure}
\epsfig{figure=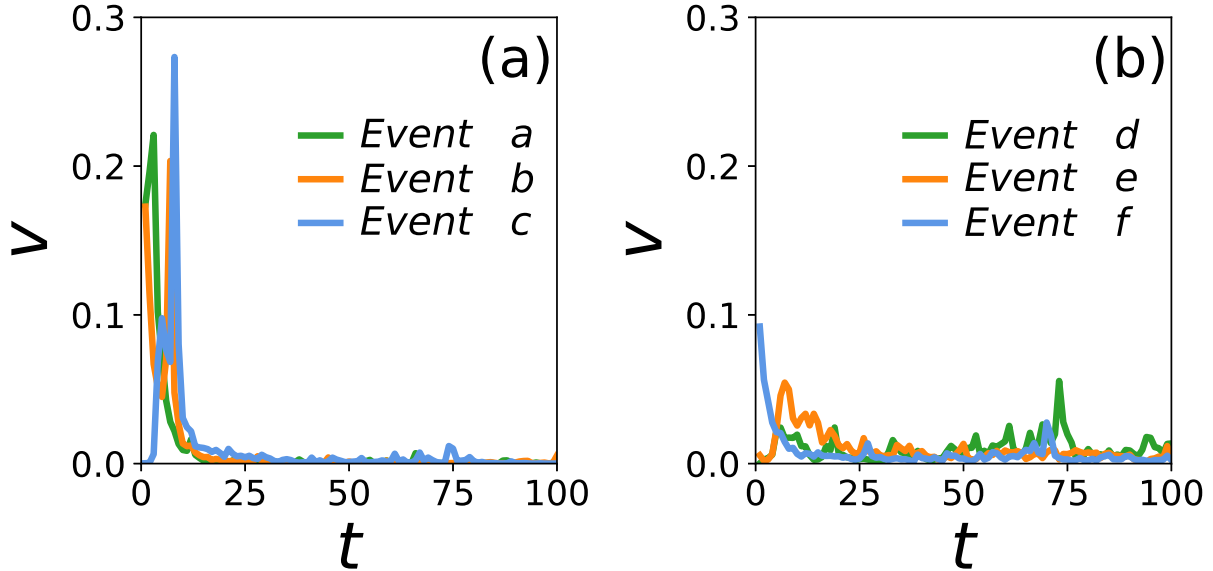,width=1.\linewidth}
\caption{(color online). (a) and (b) represent the time evolution of the spreading speeds $V$ on each day
for sensitive and insensitive events, respectively.}
\label{Fig:speed}
\end{figure}

Then, we analyze the increasing rate of messages (i.e., spreading speed) talking about a specific event on each
day, i.e. $V=\frac{C(t+\Delta t)-C(t)}{\Delta t}$, where $\Delta t$ is chosen as $\Delta t=1$ in this work.
Fig. \ref{Fig:speed} shows the time evolution of the spreading speed $V$ for different events on each day.
Comparing Fig. \ref{Fig:speed}(a) with \ref{Fig:speed}(b), it can be seen that the spreading speeds $V$ of
sensitive and insensitive events are different on each day. For the sensitive events, $V$ are larger in the first
$10$ days (Fig. \ref{Fig:speed}(a)), indicating that the events spread rapidly. However, for the insensitive ones,
their $V$ are much smaller than the cases of Fig. \ref{Fig:speed}(a), implying that they spread slowly. Therefore,
both the fast and slow spreading patterns are possible in the early stage of information spreading, depending on
whether the events are natural disaster or human related.

\section{A model for the spreading of sudden events on social networks}
\begin{figure}
\epsfig{figure=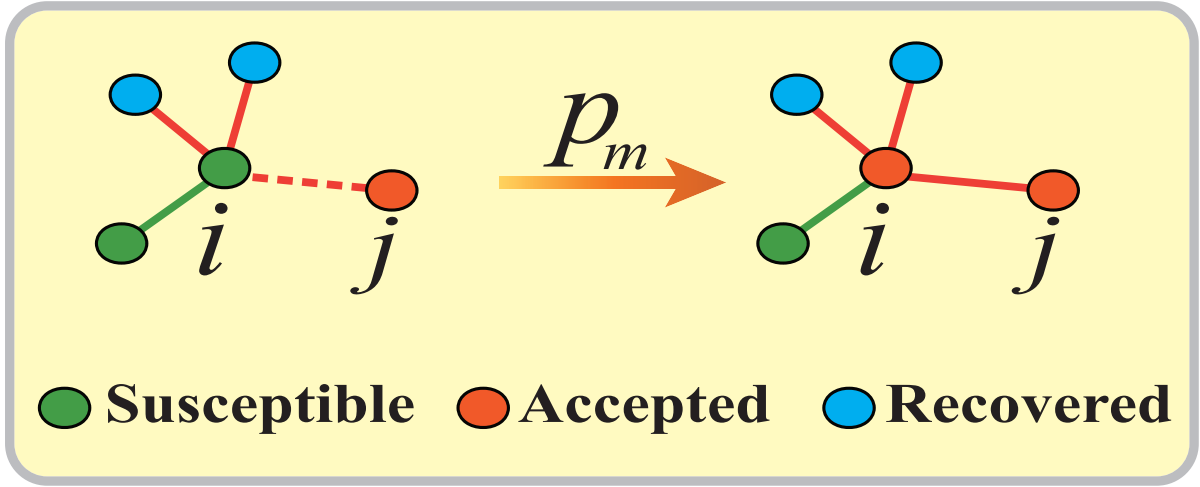,width=1.\linewidth} \caption{(color online).
Sketch of the Susceptible-Accepted-Recovered (SAR) model on a complex network, where a red solid line
denotes that the information has been transmitted successfully through it previously. At time $t$, the
susceptible node $i$ may receive the information about an event from an accepted node $j$ with probability
$\beta$ (marked with a red dashed line). Once the node $i$ receives the information successfully from
the neighbor $j$, the cumulative number $m$ of received information at the node $i$ will increase $1$.
Assuming that the node $i$ has received a new piece of information from the node $j$ at time $t$, the
cumulative number $m$ will be $3$ in this example and the node $j$ will not transmit the same information
to the node $i$ any more. Then, the accepted probability will be given by Eq. (\ref{function2}).
}
\label{Fig:sketch}
\end{figure}

To better understand the phenomena of both the fast and slow spreading patterns of sudden events on social networks,
a suitable model is needed. We here introduce such a model of information spreading on complex networks by considering
an uncorrelated network with $N$ nodes, $E$ links and degree distribution $P(k)$, where nodes represent individuals
of population and the spreading process occurs only between the neighboring nodes through links. Fig. \ref{Fig:sketch}
shows the schematic figure of this model. At each time step, a node can take only one of the three states: (i)
Susceptible: the node has not received the information about the event yet or has received the information but
hesitate to accept it; (ii) Accepted: the node accepts the information and transmits it to its neighbors; (iii)
Recovered: the node loses interest to the information and will not spread it any more. Thus, this
Susceptible-Accepted-Recovered (SAR) model is similar to the SIR (Susceptible-Infected-Refractory) model in
epidemiology.

The information spreading process can be described as follows:

(i) To initiate an event, a fraction $\rho_0$ of nodes are random uniformly chosen from the considered network as
seeds (accepted state) to spread the first piece of information. All other nodes are in the susceptible state.

(ii) At each time step $t$, every accepted node will post the information and propagate it to each of its
neighbors independently, with a transmission probability $\beta$. Once the transmission is successfully reached a
neighbor, the cumulative number $m$ of received information will be increased by one for this neighbor. In our
model, as people rarely transmit the same information to one person once and once again, an edge that has
transmitted the information successfully will never transmit the same information again, i.e., non-redundant
information transmission.

(iii) At a time step $t$, the probability for a susceptible node to accept an information is $p_m$ (see
Eq. (\ref{function2})) if it receives the information at least once at the $t$-th time step and has received it
$m$ times until time $t$. At the same time step, each accepted node will lose interest in transmitting the
information and becomes recovered with probability $\mu$.

(iv) The steps are repeated until all accepted nodes have become recovered.

Now, the key point is how to define the accepted probability $p_m$. Inspired by our previous work in
Ref \cite{Zheng:2013}, we adopt the accepted probability $p_m$ as follows. When a node receives the information
at the first time, it will accept the information with probability $p_1=\lambda$, where $\lambda$ is the
information sensitivity reflecting the sensitive intensity of information for an event. Larger $\lambda$ means
that the information is more sensitive and individuals are likely to accept the information. When a node receives
the information twice or three times, it will accept the information with a probability $p_2$ or $p_3$,
respectively. In our model, the accepted probability with different received times is defined as following:
\begin{eqnarray}
p_1 &=& \lambda\nonumber,\\
p_2 &=& p_1+\eta\times(1-p_1) \nonumber,\\
p_3 &=& p_2+\eta\times(1-p_2)\nonumber, \\
&\vdots& \nonumber\\
p_m &=& p_{m-1}+\eta\times(1-p_{m-1}),
\label{function1}
\end{eqnarray}
where $\eta\in[0,1]$ is the social reinforcement strength. A larger $\eta$ means that the redundant information
will have stronger influence on nodes. The iterative Eq. (\ref{function1}) indicates that if a node has received
the information $m$ times, the accepted probability will increase $\eta\times(1- p_{m-1}$), comparing with
$p_{m-1}$. The increase can be considered as an increment of spreading converted from disapproving
probability $1-p_{m-1}$ under the effect of social reinforcement. In sum, the Eqs. (\ref{function1}) can be
simplified into
\begin{eqnarray}
p_m= 1-(1-\lambda)(1-\eta)^{m-1}, 0\leq \eta \leq1, m\geq 1.
\label{function2}
\end{eqnarray}
It is found that by Eq. (\ref{function2}), the simulation results can well explain the results of Centola's
experiments on behavior spreading and some former studies on information spreading in different parameter spaces
\cite{Zheng:2013,Centola:2010}. We are wondering whether this model can be used to reproduce the spreading
patterns of real data in Fig. \ref{Fig:data}. It is worth noting that our model has three key parameters: the
transmission probability $\beta$, the information sensitivity $\lambda$ and the social reinforcement strength
$\eta$. Meanwhile, this model emphasizes the effect of the information sensitivity, social reinforcement and
non-redundant information memory, which make the information spreading processes be non-Markovian.

\section{Results}
\subsection{Reproduce the spreading patterns of the six typical events in Sina Weibo}
\begin{figure}
\epsfig{figure=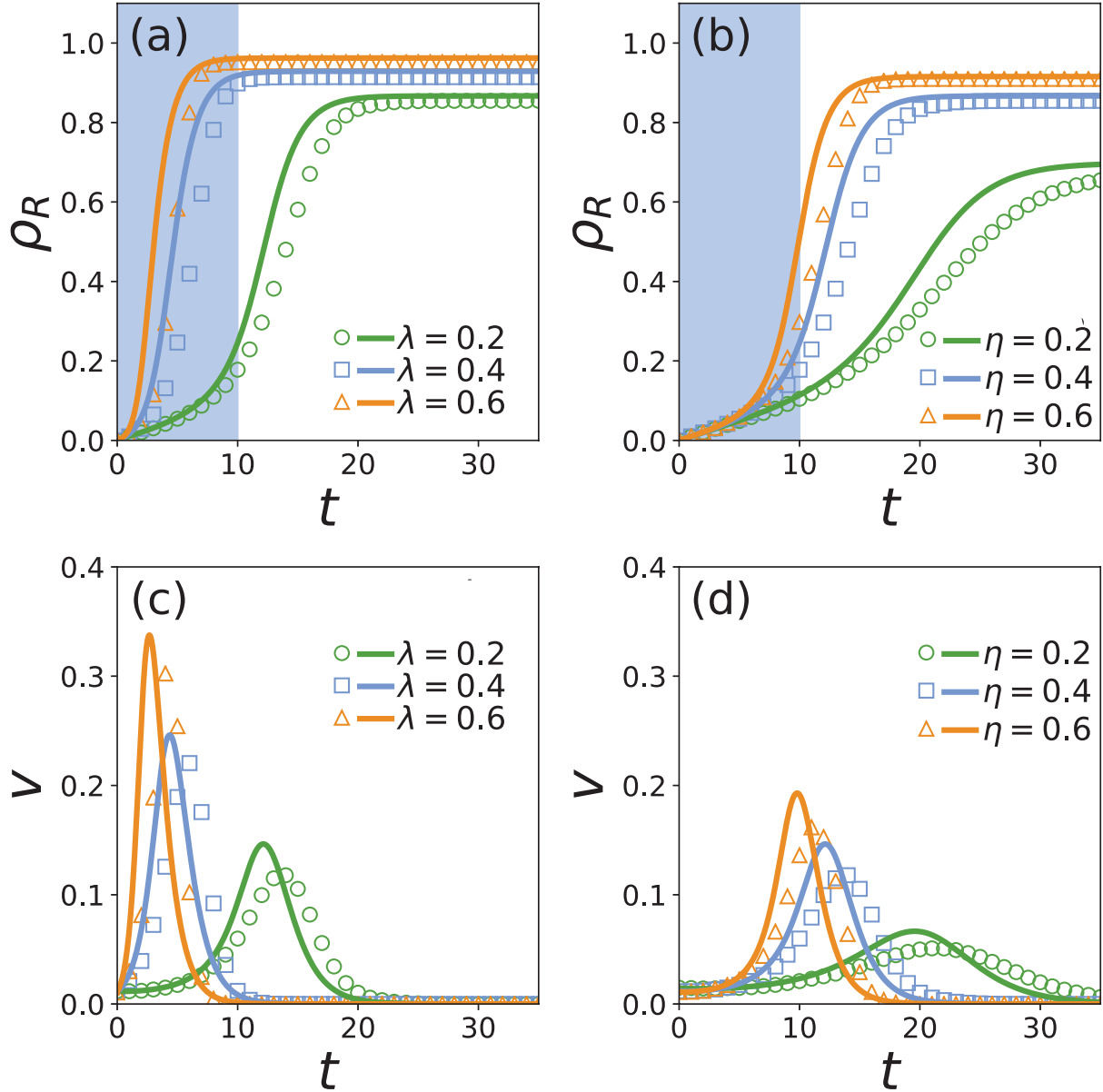,width=1.\linewidth} \caption{(color online).
(a) and (b) represent the fraction of recovered nodes $\rho_R$ as a function of time $t$ with different
information sensitivities $\lambda$ and different social reinforcements $\eta$, respectively. Light blue areas
represent the spreading range within $10$ time steps in the spreading process.
(c) and (d) represent the information spreading speed $V$ as a function of time $t$ corresponding to (a) and
(b), respectively. The symbols represent the simulated results and the lines are the corresponding theoretical
results. The parameters are fixed as $\eta=0.4$ in (a) and (c) and $\lambda=0.4$ in (b) and (d), respectively.
Other parameters are set as $N=10\,000$, $\beta=0.8$, $\rho_0=0.01$, and $\mu=1.0$. All the results are
averaged over 100 independent realizations.
}
\label{Fig:timeseries}
\end{figure}

In numerical simulations, we choose the network as the Erd\H{o}s-R\'{e}nyi (ER) random network with
size $N=10\,000$ and average degree $\langle k \rangle=6$ and study the information spreading process on it. We
let $\rho_0=0.01$ and $\mu=1.0$ in this paper. We let $\rho_R(t)$ denote the fraction of recovered nodes at
time $t$ in the spreading process, which corresponds to the quantity $C$ in the empirical data. In stationary state,
$\rho_R(t)$ represents the range of spreading. Fig. \ref{Fig:timeseries}(a) shows the time evolution of $\rho_R$
for $\eta=0.4$ where the ``circles", ``squares" and ``triangles" represent the cases of $\lambda=0.2, 0.4$ and
$0.6$, respectively. We see that when $\lambda$ is large, $\rho_R$ increases sharply
in the first $10$ time steps and shows the similar patterns as that in Figs. \ref{Fig:data}(a)-(c) (see the light
blue shadowed areas). The incremental value $\Delta$ of recovered nodes within the first $10$ time steps (i.e.,
$\Delta=\rho_R(10)-\rho_R(0)$) can reach $89.83\%$ and $95.15\%$ with $\lambda=0.4$ and $0.6$, respectively,
which confirm the characteristic feature of rapidly spreading again. When $\lambda$ is small, such as
$\lambda=0.2$, $\rho_R$ increases slowly and its spreading pattern is similar to that of Figs. \ref{Fig:data}(d)-(f).
Its $\Delta$ is only $17.79\%$ in the first $10$ time steps (see the light blue shadowed areas). Thus, the model
can show the main patterns of both the fast and slow spreading in the real Weibo data. These results can be also
theoretically predicted by the edge-based compartmental theory (see the next section for details). The solid lines in
Fig. \ref{Fig:timeseries}(a) represent the theoretical results from Eq. (\ref{f14}). We see that the theoretical
results are consistent with the numerical simulations very well.

To compare with the spreading speed $V$ in Fig. \ref{Fig:speed}, we here redefine
$V=\frac{\rho_R(t+\Delta t)-\rho_R(t)}{\Delta t}$ and set $\Delta t=1$ as in the empirical analyses. Notice
that the recovered probability $\mu$ is set to be unity in this work. When $V=0$, no accepted nodes will be
turned into the recovered state, implying that there is no accepted nodes in the system. Thus, the spreading
process will be ended once $V=0$ and the spreading range will reach its maximum.
Fig. \ref{Fig:timeseries}(c) shows the time evolution of $V$, corresponding to Fig. \ref{Fig:timeseries}(a).
It is easy to see that when $\lambda$ is large, the value of $V$ is also large and the peak of $V$ is
located in the first $10$ time steps, indicating that the information spreads rapidly in the early stage.
This result is consistent with the empirical observations in Fig. \ref{Fig:speed}(a). While for the case of
small $\lambda$, i.e. $\lambda=0.2$, $V$ will be smaller than that in the case of large $\lambda$, confirming
that its spreading of information is slow in the early stage. From Fig. \ref{Fig:timeseries}(c) we also
notice that the peak of $V$ is out of the first $10$ time steps, indicating that the occurrence of outbreak
has been delayed. This result is consistent with Fig. \ref{Fig:speed}(b). In addition, the obtained results
have been confirmed by the theoretical results (see the next section for details). The solid lines in
Fig. \ref{Fig:timeseries}(c) represent the theoretical results from Eq. (\ref{f14}). Once again, we see that
the theoretical results are consistent with the numerical simulations very well. Therefore, it can be concluded
that the model can show the main patterns of both the fast and slow spreading in real data.

Besides the effect of information sensitivity, another important problem is how the social reinforcement $\eta$
influences the range and speed of information spreading. To answer this question, we plot the time evolution
of the fraction of recovered nodes $\rho_R$ with fixed $\lambda=0.4$ in Fig. \ref{Fig:timeseries}(b) where the
``circles", ``squares" and ``triangles" represent the cases of $\eta=0.2, 0.4$ and $0.6$, respectively. We see
that the difference between the values of $\rho_R$ with different $\eta$ is insensitive at $t<8$. After that,
with the further increasing of $t$, the difference will become larger and larger, implying that the influence
of social reinforcement takes effect mainly after the early stage. In this stage of $t>8$, an individual has
more chance to receive multiple information and thus the accepted probability is increased by the social
reinforcement. This feature is also reflected in the spreading speed $V$, see Fig. \ref{Fig:timeseries}(d),
where a larger $\eta$ will accelerate the spreading of information. Thus, the social reinforcement is another
key factor to influence the information spreading.

\subsection{Effects of dynamical parameters}
\begin{figure}
\epsfig{figure=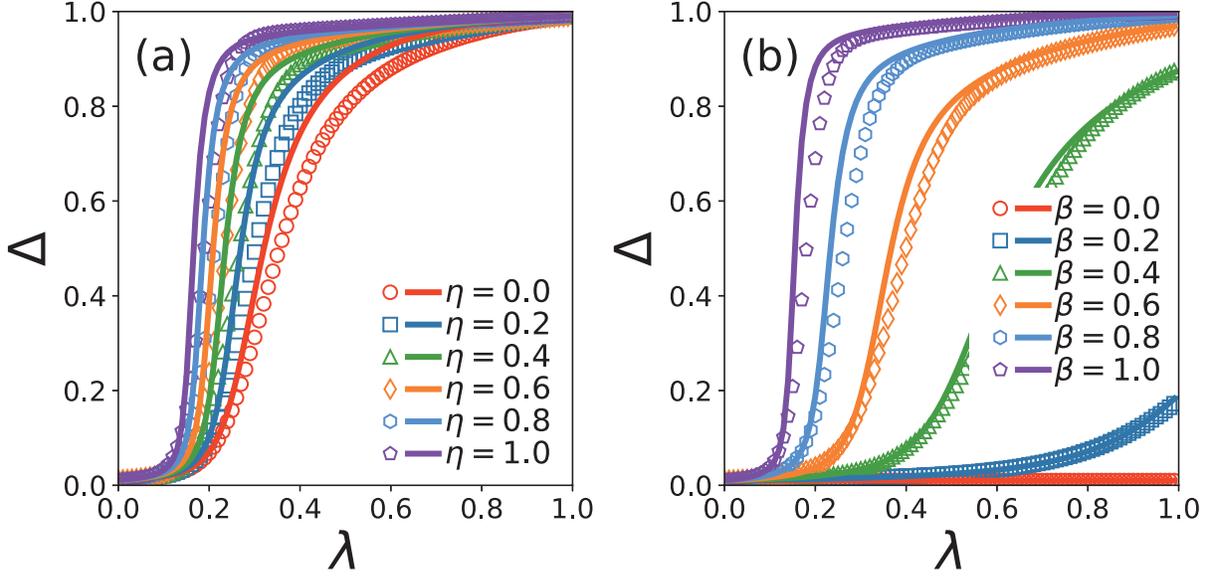,width=1.\linewidth} \caption{(color online).
(a) and (b) represent the dependence of $\Delta$ within the first $10$ time steps on $\lambda$ for different
$\eta$ and $\beta$, respectively.
The symbols represent the simulated results and the lines are the corresponding theoretical results. The parameters
are fixed as $\beta=0.8$ in (a) and $\eta=0.4$ in (b), respectively. Other parameters are set as the same as in
Fig. \ref{Fig:timeseries}.
}
\label{Fig:simDelta}
\end{figure}

To quantitatively and deeply understand the effects of $\lambda$ and $\eta$ in the early stage \cite{Liu:2015},
we investigate their influences on the incremental rate $\Delta$ of recovered nodes as in the empirical analysis.
Fig. \ref{Fig:simDelta} shows the the dependence of $\Delta$ on $\lambda$ for different $\eta$ and $\beta$,
respectively. We see that $\Delta$ increases gradually with $\lambda$ for each fixed $\eta$. When $\eta$
gradually increases from $0$ to $1$, the increase of $\Delta$ will change from slowly to sharply. Therefore,
both the information sensitivity $\lambda$ and social reinforcement $\eta$ will increase the value of $\Delta$
and thus accelerate the information spreading.

Except the two parameters $\lambda$ and $\eta$, we find that the transmission probability $\beta$ also plays a
key role on $\Delta$. Fig. \ref{Fig:simDelta}(b) shows the results where $\Delta$ increase slowly when $\beta$
is small but rapidly when $\beta$ is large. This result can be understood as follows. As a larger $\beta$ will
make a node have a larger probability to receive information from its neighbors, the redundant information will
increase the accepted probability $p_m$ and thus result in the fast information spreading. This conclusion will
be confirmed by theoretical results in the next section.

\begin{figure}
\epsfig{figure=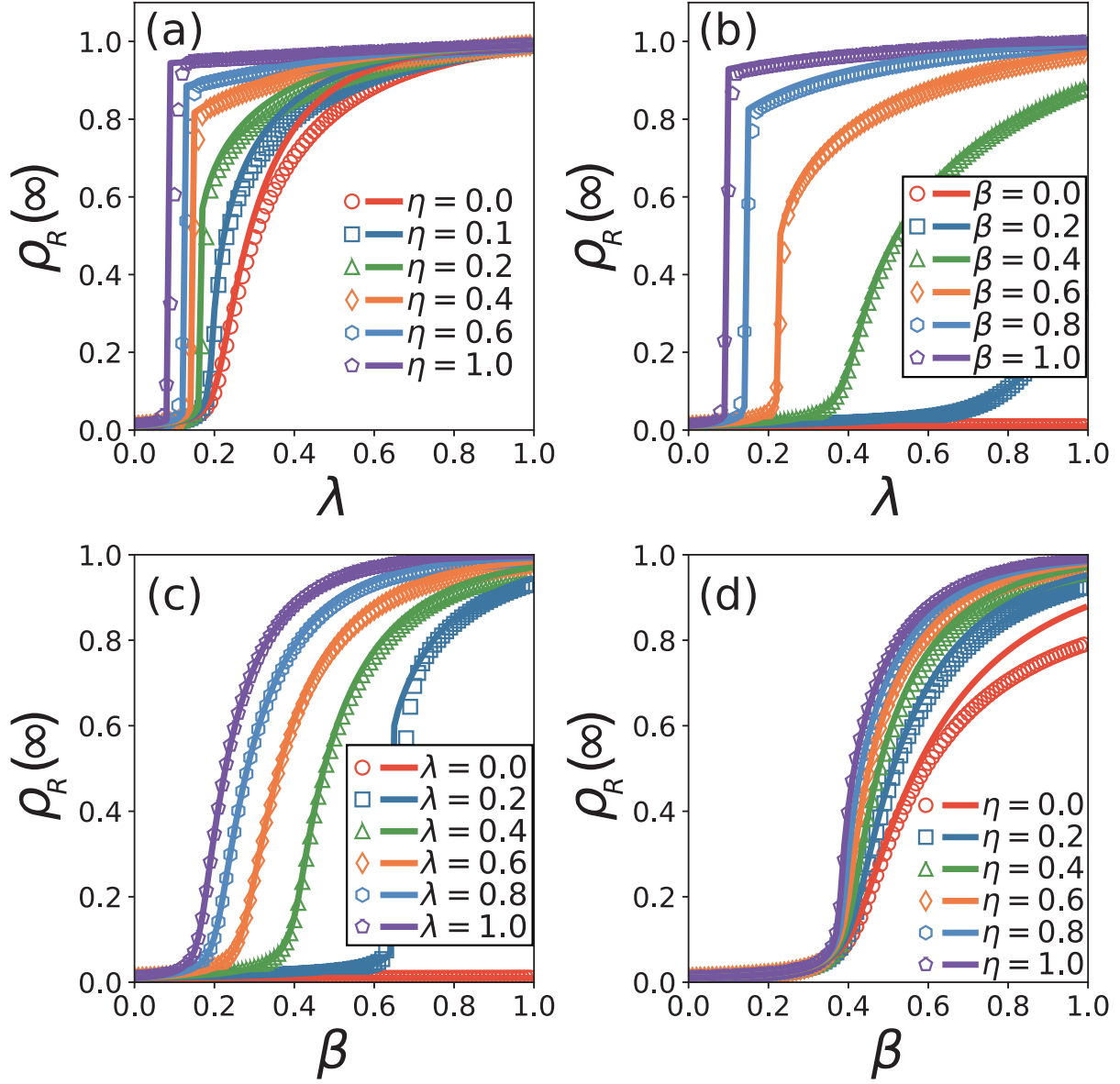,width=1.\linewidth} \caption{(color online).
(a) and (b): The final accepted size $\rho_R(\infty)$ versus the information sensitivity $\lambda$ for different
$\eta$ in (a) and different $\beta$ in (b).
(c) and (d): The final accepted size $\rho_R(\infty)$ versus the transmission probability $\beta$ for different
$\lambda$ in (c) and different $\eta$ in (d).
The symbols represent the simulated results and the lines are the corresponding theoretical
results. The parameters are fixed as $\beta=0.8$ in (a), $\eta=0.4$ in (b), $\eta=0.4$ in (c) and $\lambda=0.4$
in (d), respectively. The other parameters are the same as in Fig. \ref{Fig:timeseries}.
}
\label{Fig:Finalsize}
\end{figure}
Another key quantity of spreading dynamics is the final size of accepted nodes, denoted by $\rho_R(\infty)$. A
larger $\rho_R(\infty)$ implies a larger spreading range at the final state. Fig. \ref{Fig:Finalsize}(a) shows
the dependence of $\rho_R(\infty)$ on $\lambda$ for different $\eta$. We surprisingly find that $\rho_R(\infty)$
increases continuously with $\lambda$ when $\eta$ is small but discontinuously when $\eta$ is large. Let
$\lambda_c$ represent the critical value for $\rho_R(\infty)$ to change from zero to nonzero. The outbreak
transition on $\lambda_c$ will be continuous for a small $\eta$ but discontinuous for a large $\eta$. As a small $\eta$ will
reduce the accepted probability $p_m$, the node will not accept the information until it receives the information
multiple times (i.e., $m$ should be large). Thus, when the social reinforcement is small, a node is not likely to
accept the information and thus the spreading range $\rho_R(\infty)$ is small.

Moreover, the information transmission rate $\beta$ also has significant influence on the final accepted size
$\rho_R(\infty)$. Fig. \ref{Fig:Finalsize}(b) shows the dependence of $\rho_R(\infty)$ on $\lambda$ for different
$\beta$. It is easy to see that when $\beta=0$, the information can not be spread out for all the $\lambda$. This
is a trivial case as individuals can not receive any information. Once $\beta$ is not zero (i.e., $\beta=0.2$),
$\rho_R(\infty)$ will increase continuously with $\lambda$. When the transmission probability is larger
(i.e., $\beta=0.6$), the system also shows a discontinuous phase transition at the critical value $\lambda_c$.
As a large transmission probability is preferred to receive the information for individuals, it promotes the
information spreading. This result has been confirmed by Eq. (\ref{f16}) of the theory, see the lines in
Fig. \ref{Fig:Finalsize}(b).

To understand the effect of transmission probability $\beta$ deeply, Fig. \ref{Fig:Finalsize}(c) and (d) show the
dependence of $\rho_R(\infty)$ on the transmission probability $\beta$ for different $\lambda$ and $\eta$,
respectively. From Fig. \ref{Fig:Finalsize}(c), we have $\rho_R(\infty)=0$ when $\lambda=0$, indicating that no one
have interest to spread information. When $\lambda>0$, there is a critical point $\beta_c$ where we have
$\rho_R(\infty)=0$ for $\beta<\beta_c$ and $\rho_R(\infty)>0$ for $\beta>\beta_c$. Very interesting, we find that
$\beta_c$ will decrease with the increase of $\lambda$, confirming that a large $\lambda$ will promote the
information spreading.

However, for the social reinforcement $\eta$, the spreading is different from the case of $\lambda$.
Fig. \ref{Fig:Finalsize}(d)) shows the results. It is easy to see that the critical value $\beta_c$ is insensitive
to $\eta$. The reason is that a node is unlikely to receive the information multiple times in early stage, thus
the influence of social reinforcement on $\rho_R(\infty)$ becomes less important in the stage.

From the above discussions, we conclude that $\beta$, $\lambda$ and $\eta$ are the three key parameters to
significantly influence $\rho_R(\infty)$ and the phase transition. Thus, they are of significance on the
information spreading on real social networks. The observed discontinuous phase transition may explain the mechanism
why the information can suddenly and sometimes unexpectedly catch on \cite{Gardenes:2016}.

\section{A theoretical analysis based on edge-based compartmental theory}
To explain the information spreading patterns of the above numerical results, we here make a theoretical analysis. We
apply the edge-based compartmental theory on complex networks by following the methods and tools introduced
in Refs. \cite{Wang:2015,Volz:2008,Miller:2011,Shu:2016,Miller:2012,Miller:2013a,Miller:2014}. We let $\rho_S(t)$,
$\rho_A(t)$, and $\rho_R(t)$ be the densities of the Susceptible, Accepted, and Recovered nodes at time $t$,
respectively. The spreading process will be ended when $t\rightarrow\infty$ and thus $\rho_R(\infty)$ represent the
final fraction of accepted nodes.

We use a variable $\theta(t)$ to denote the probability that a node $v$ has not transmitted the information to the
node $u$ along a randomly chosen edge by time $t$. For an uncorrelated, large and sparse network, the probability
that a randomly chosen node $u$ of degree $k$ has received the information from distinct neighbors $m$ times at
time $t$ is
\begin{eqnarray}
\phi_m(k,\theta(t))={k \choose m}\theta(t)^{k-m}[1-\theta(t)]^{m}.
\label{f3}
\end{eqnarray}
Notice that a node with degree $k$ has the probability $1-\rho_0$ to be not one of the initial seeds. The
probability that an arbitrary node has not accepted the information after receiving such information $m$ times
is $\prod_{j=1}^{m}(1-p_j)=(1-\lambda)^{\Sigma_{j=1}^m j}(1-\eta)^{\Sigma_{j=1}^mj-1}$. Then, the probability
that a susceptible node $u$ with degree $k$ has received the information $m$ times and still does not accept it
by time $t$ is $\phi_m(k,\theta(t))(1-\lambda)^{\Sigma_{j=1}^m j}(1-\eta)^{\Sigma_{j=1}^mj-1}$.
Combining the initial seeds and summing over all possible values of $m$, we obtain the probability that the node
$u$ is still in the susceptible state at time $t$ as
\begin{eqnarray}
S(k,t)&=&(1-\rho_0)\sum\limits_{m=0}^{k}\phi_m(k,\theta(t)) \nonumber \\
&&\times(1-\lambda)^{\sum_{j=1}^m j}(1-\eta)^{\sum_{j=1}^mj-1}.
\label{f4}
\end{eqnarray}
Averaging over all $k$, the density of susceptible nodes (i.e., the probability of a randomly chosen individual
is in the susceptible state) at time $t$ is given by
\begin{eqnarray}
\rho_S(t)=\sum\limits_{k=0}^{\infty}P(k)S(k,t).
\label{f5}
\end{eqnarray}

Since a neighbor $v$ of node $u$ may be susceptible, infected, or recovered, $\theta(t)$ can be expressed as
\begin{equation}
\theta(t)=\Phi^S(t)+\Phi^A(t)+\Phi^R(t).
\label{f6}
\end{equation}
where $\Phi^S(t),\Phi^A(t),\Phi^R(t)$ is the probability that the neighbor $v$ is in the susceptible, accepted,
recovery state, respectively, and has not transmitted the information to node $u$ through their connections. Once
these three parameters are derived, we will get the density of susceptible nodes at time $t$ by substituting
them into Eq. (\ref{f4}) and then into Eq. (\ref{f5}). To this purpose, in the following, we will focus on how
to derive them.

To find $\Phi^S(t)$, we consider a randomly selected node $u$ with degree $k$, and assume that this node is in
the cavity state, which means that it cannot transmit any information to its neighbors $v$ but can receive
such information from its neighbors. In this case, the neighbor $v$ can only get information from its other
neighbors except the node $u$. If a neighboring node $v$ of $u$ has degree $k'$, the probability that node
$v$ has received $m$ pieces of the information at time $t$ will be
\begin{eqnarray}
\psi_m(k',\theta(t))={{k'-1} \choose m}\theta(t)^{k'-m-1}[1-\theta(t)]^{m}.
\label{f7}
\end{eqnarray}
Similar to Eq. (\ref{f4}), individual $v$ will still stay in the susceptible state by time $t$ with the
probability
\begin{eqnarray}
\Theta(k',\theta(t))&=&(1-\rho_0)\sum\limits_{m=0}^{k'-1}\psi_m(k',\theta(t)) \nonumber \\
&&\times(1-\lambda)^{\sum_{j=1}^m j}(1-\eta)^{\sum_{j=1}^mj-1}.
\label{f8}
\end{eqnarray}

For uncorrelated networks, the probability that one edge from node $u$ connects with a node with
degree $k'$ is $k'P(k')/\langle k\rangle$, where $\langle k\rangle$ is the mean degree of the network. Summing
over all possible $k'$, we obtain the probability that $u$ connects to a susceptible node by time $t$ as
\begin{eqnarray}
\Phi_S(t)=\frac{\sum_{k'}k'P(k')\Theta(k',\theta(t))}{\langle k\rangle}.
\label{f9}
\end{eqnarray}

According to the information spreading process as described in Sec.II, the growth of $\Phi^R(t)$ includes two
consecutive events: firstly, an accepted neighbor has not transmitted the information successfully to node $u$
with probability $1-\beta$; secondly, the accepted neighbor has been recovered with probability $\mu$. Combining
these two events, the $\Phi^I(t)$ to $\Phi^R(t)$ flux is $\mu(1-\beta)\Phi^I(t)$. Thus, one gets
\begin{equation}
\frac{d\Phi^R(t)}{dt}= \mu(1-\beta)\Phi^A(t).
\label{f10}
\end{equation}

Once the accepted neighbor $v$ transmits the information to $u$ successfully (with probability $\beta$), the
$\Phi^A(t)$ to $1-\theta(t)$ flux will be $\beta\Phi^A(t)$, which means
\begin{eqnarray}
\frac{d(1-\theta(t))}{dt}=\beta\Phi^A(t). \nonumber
\end{eqnarray}
That is
\begin{equation}
\frac{d\theta(t)}{dt}=-\beta\Phi^A(t).
\label{f11}
\end{equation}

Combining Eqs. (\ref{f10}) and (\ref{f11}) and considering (as initial conditions) $\theta(0)=1$ and
$\Phi^R(0)=0$, one obtains
\begin{eqnarray}
\Phi_R(t)=\frac{\mu[1-\theta(t)](1-\beta)}{\beta}.
\label{f12}
\end{eqnarray}

Substituting Eqs. (\ref{f9}) and (\ref{f12}) into Eq.(\ref{f6}), we get an expression for $\Phi^A(t)$ in terms
of $\theta(t)$. Then, one can rewrite Eq. (\ref{f11}) as
\begin{eqnarray}
\frac{d\theta(t)}{dt}&=&-\beta \left[\theta(t)-\frac{\sum_{k'}k'P(k')\Theta(k',\theta(t))}{\langle k\rangle} \right] \nonumber\\
&&+\mu[1-\theta(t)](1-\beta).
\label{f13}
\end{eqnarray}

With $\theta(t)$ on hand, the equation of the system comes out to be
\begin{eqnarray}
\frac{d\rho_R(t)}{dt}&=&\mu \rho_A(t) \nonumber, \\
\rho_S(t)&=&\sum\limits_{k=0}^{\infty}P(k)S(k,t) \nonumber, \\
\rho_A(t)&=&1-\rho_S(t)-\rho_R(t).
\label{f14}
\end{eqnarray}
Eq. (\ref{f14}) is the main theoretical result which gives the densities of $\rho_S(t), \rho_A(t)$ and $\rho_R(t)$ at
time $t$.

Furthermore, we can obtain the final accepted size $\rho_R(\infty)$ in the steady state (i.e., the final fraction
of nodes that have accepted the information). By setting $t\rightarrow \infty$ and $\frac{d\theta(t)}{dt}=0$ in
Eq.(\ref{f13}), we get
\begin{eqnarray}
\theta(\infty)&=&\frac{\sum_{k'}k'P(k')\Theta(k',\theta(\infty))}{\langle k\rangle} \nonumber\\
&&+\frac{\mu[1-\theta(\infty)](1-\beta)}{\beta}.
\label{f15}
\end{eqnarray}
Substituting $\theta(\infty)$ into Eqs.(\ref{f3})$-$(\ref{f5}), we can calculate the value of $\rho_S(\infty)$,
and then the final accepted size can be obtained as
\begin{eqnarray}
\rho_R(\infty)=1- \rho_S(\infty).
\label{f16}
\end{eqnarray}

Instead of getting the analytic solutions of Eqs. (\ref{f14}) and (\ref{f16}), we solve them by numerical integration.
By this way, we can obtain the solutions of Eq. (\ref{f14}) in Figs. \ref{Fig:timeseries} and \ref{Fig:simDelta}, which
show the similar pattern to the simulation and empirical results. The pattern of fast spreading is likely to appear for
a large information sensitivity while the pattern of slow spreading tends to be triggered for a small information
sensitivity. In addition, according to Eq. (\ref{f16}), we obtain the theoretical curves in Fig. \ref{Fig:Finalsize},
which are consistent with the numerical results very well and thus confirm the effects of $\beta$, $\lambda$ and $\eta$
and the phase transition.

\section{Conclusions and Discussions}

The information spreading on networks is a very hot topic in the field of complex network in recent years, which
focuses mainly on how the spreading is influenced by the network structure and other significant properties. However,
to our knowledge, it does not take into account the effects of both the information sensitivity and social reinforcement.
The former reflects the effect of event attribute indirectly and the latter indicates the fact that accepting a piece of
information requires verification of its credibility and legitimacy, both being the key ingredients in information
dynamics. At the same time, little attention has been paid to the study of combining the real data and theoretical model.
Thanks to the fast development of database technology and computational power, we can obtain some spreading data of the
typical events from Sina Weibo. With the supports of these data, we can go deeply to understand the impacts of both the
information sensitivity and social reinforcement.

In summary, we have proposed a SAR model to describe the information spreading patterns of six typical events in Sina
Weibo, which includes two essential properties of the information spreading, i.e. information sensitivity and social
reinforcement. By both numerical simulations and theoretical analysis we show that the information spreading can be
either extremely fast or very slow, which agrees well with empirical data. The spreading patterns may be influenced by
either the information sensitivity or social reinforcement. Especially, when the strength of the social reinforcement is
large, an explosive phase transition can be expected in the parameter space. These findings may provide an explanation
for the extremely fast spreading of modern fashion such as the news, rumours, products etc.

The main contributions of this work include the discovery of both the fast and slow spreading patterns from the data
of Sina Weibo, and a qualitative and quantitative understanding of the phenomena by the SAR model. However, many
challenges still remain. For example, more real data of information spreading are needed to further test the validity
of the model. Moreover, the effects of network structure remain to be studied on information spreading dynamics, such
as the degree heterogeneity \cite{Newman:2003}, clustering \cite{Serrano:2006,Newman:2009}, community
\cite{Girvan:2002,Fortunato:2010,Gong:2013}, and core periphery \cite{Borgatti:2000,Holme:2005,LiuY:2015,Verma:2016}
etc. Finally, the study may be extended to more realistic networks such as multi-layer networks
\cite{Boccaletti:2014,Gu:2011,Zheng:2017}, temporal networks \cite{Holme:2012} and so on.

This work was partially supported by the NNSF of China under Grant Nos. 11675056, 11375066 and 11505114,
and the Program for Professor of Special Appointment (Orientational Scholar) at Shanghai Institutions of Higher
Learning under Grants No. QD201.

\end{document}